\pdfoutput=1 
\documentclass[letterpaper, 10 pt, conference]{ieeeconf}  
\usepackage{cite}
\usepackage{amsmath,amssymb,amsfonts}
\usepackage{algorithmic}
\usepackage{graphicx}
\usepackage{textcomp}
\usepackage{color,soul}
\usepackage{caption}
\usepackage{subcaption}
\usepackage{booktabs} 
\usepackage{makecell} 
\usepackage[flushleft]{threeparttable}     

\IEEEoverridecommandlockouts                              

\title{\LARGE \bf
Workload Assessment of Human-Machine Interface: A Simulator Study with Psychophysiological Measures 
}

\author{Yuan-Cheng Liu$^{1}$, Nikol Figalová$^{2}$, Jürgen Pichen$^{3}$, Philipp Hock$^{4}$, Martin Baumann$^{5}$, and Klaus Bengler$^{6}$
\thanks{$^{1}$Yuan-Cheng Liu is with the Chair of Ergonomics, School of Engineering and Design,
        Technical University of Munich, 85748 Garching, Germany
        {\tt\small yuancheng.liu@tum.de}}%
\thanks{$^{2}$Nikol Figalová is with the Department of Clinical and Health Psychology,
        Ulm University, 89069 Ulm, Germany
        {\tt\small nikol.figalova@uni-ulm.de}}%
\thanks{$^{3}$Jürgen Pichen is with the Department of Human Factors,
        Ulm University, 89069 Ulm, Germany
        {\tt\small juergen.pichen@uni-ulm.de}}%
\thanks{$^{4}$Philipp Hock is with the Department of Human Factors,
        Ulm University, 89069 Ulm, Germany
        {\tt\small philipp.hock@uni-ulm.de}}%
\thanks{$^{5}$Martin Baumann is with the Department of Human Factors,
        Ulm University, 89069 Ulm, Germany
        {\tt\small martin.baumann@uni-ulm.de}}%
\thanks{$^{1}$Klaus Bengle is with the Chair of Ergonomics, School of Engineering and Design,
        Technical University of Munich, 85748 Garching, Germany
        {\tt\small bengler@tum.de}}%
}

\begin{document}

\maketitle
\thispagestyle{empty}
\pagestyle{empty}

\begin{abstract}

Human-machine Interface (HMI) is critical for safety during automated driving, as it serves as the only media between the automated system and human users. To enable a transparent HMI, we first need to know how to evaluate it. However, most of the assessment methods used for HMI designs are subjective and thus not efficient. To bridge the gap, an objective and standardized HMI assessment method is needed, and the first step is to find an objective method for workload measurement for this context. In this study, two psychophysiological measures, electrocardiography (ECG) and electrodermal activity (EDA), were evaluated for their effectiveness in finding differences in mental workload among different HMI designs in a simulator study. Three HMI designs were developed and used. Results showed that both workload measures were able to identify significant differences in objective mental workload when interacting with in-vehicle HMIs. As a first step toward a standardized assessment method, the results could be used as a firm ground for future studies.

\end{abstract}

\section{INTRODUCTION}

Human-machine Interface (HMI) of Automated Driving Systems (ADS) is a critical component that aims to facilitate an intuitive way for humans to interact with automation \cite{bengler2020hmi}. However, depending on the level of automation, the roles and responsibilities of human users may be constantly changing. Thus, the design of the HMI plays a crucial part in enabling a safe and efficient transition between the roles by providing critical and understandable information and making the automated system transparent.

When evaluating HMIs, existing methods are mostly based on subjective questionnaires \cite{richardson2018conceptual, voinescu2020utility}, making them prone to biases and difficult to standardize. To resolve the problem, we proposed the Transparency Assessment Method (TRASS) in a previous study \cite{liu2022transparency}, where the transparency toward the HMI is estimated by evaluating the actual understanding and the workload of the user during the interaction. To further apply the TRASS in a dynamic environment (e.g., in simulator or test track studies), the workload estimation method must be adapted. 

Multidisciplinary approaches have been used to measure workload in driving scenarios \cite{stapel2019automated, kim2018effectiveness, lim2018stew, matthews2019monitoring}. Subjective mental workload measures like NASA-TLX are easy to assess and come directly from participants. However, these subjective mental workload assessment methods alone are usually inaccurate and unable to operate in real-time. On the other hand, psychophysiological measures enable a more objective and continuous assessment of the mental workload. With the non-intrusive sensors, the experimental process becomes more efficient, and the resulting measurements are also more reliable.

In this paper, we present a simulator study where electrocardiography (ECG) and electrodermal activity (EDA) are used to estimate the objective workload of participants when interacting with AV HMIs. The aim of this paper is to evaluate the effectiveness of these psychophysiological measures in finding differences in mental workload when interacting with different HMI designs in simulated driving. To the best of our knowledge, this is the first study that applies these two psychophysiological measures to estimate the mental workload of different AV HMIs. The main novel contribution of this paper is that the results of these real-time and objective workload measures could be further applied for researches in high-fidelity AV environments and HMI design and evaluation processes.

\section{RELATED WORKS}

Psychophysiological measures have been widely adopted in the driving context to assess many psychological constructs, such as workload and stress, owing to their sensitivity and accuracy in detecting mental workload \cite{meng2022cognitive, lohani2019review}. Besides, they can also capture dynamic changes in workload that might be difficult to detect with subjective or behavior measures \cite{charles2019measuring}. In this study, we use ECG and EDA for their reliability in short task duration and sensitivity in continuous mental workload detection \cite{baek2015reliability, yoshida2014feasibility}.

\subsection{Electrocardiography}

Electrocardiography (ECG) is a method commonly used to capture the electrical activity of the heart. From a series of heart beat waves, heart rate (HR) and heart rate variability (HRV) could be calculated, analyzed, and used to estimate mental workload \cite{heine2017electrocardiographic, shakouri2018analysis}. HR was found to increase with the increase in the mental workload, while the HRV decreased. 

Heart period (R-R interval) could be derived from the time intervals between heart beats (R peaks). By converting the heart period (usually in milliseconds), we could obtain the heart rate (usually in beats per minute). On the other hand, HRV metrics are more versatile and could be categorized into frequency domain and time domain. In the frequency domain, methods such as low-frequency (LF) power, high-frequency (HF) power, or LF/HF ratio are often used \cite{alaimo2020aircraft}. While in the time domain, the standard deviation of R-R intervals (SDRR) and root mean square of successive differences between normal heartbeats (RMSSD) are widely adopted. However, literature shows that the RMSSD is one of the most robust workload measurements, which is also reliable in ultra-short-term analysis (measurement duration less than 5 mins) \cite{shaffer2017overview, baek2015reliability}.

\subsection{Electrodermal Activity}

Electrodermal activity (EDA) reflects changes in the electrical potential of the skin. It has been commonly used as an objective workload indicator in simulators and real driving studies \cite{yoshida2014feasibility, daviaux2020event}. Two components are usually derived from the EDA signals, which are tonic and phasic. The tonic component is the slowly changing in the electrical conductivity level, also known as skin conductance level (SCL). In comparison, the phasic component describes the event-related change that increases the magnitude of electrical conductance, also called skin conductance response (SCR). Both SCL and SCR are found sensitive to changes in mental workload. SCL was found to increase with the higher workload during real-driving scenarios \cite{mehler2019demanding}, while higher SCR values were found with higher mental demands \cite{foy2018mental}.

\section{METHODOLOGY}

\subsection{Participants}

Twenty-four participants were recruited for this study, where eight were male, 15 were female, and one was diverse. The ages ranged from 22 to 30, with mean age $= 27.92$, $SD = 4.34$. All of the participants came with valid driving licenses, and they had held them for at least three years ($M = 8.03, SD = 3.78$)

\subsection{Human-machine Interface Designs}

\begin{figure}[htbp]
     \centering
     \includegraphics[width=0.45\textwidth]{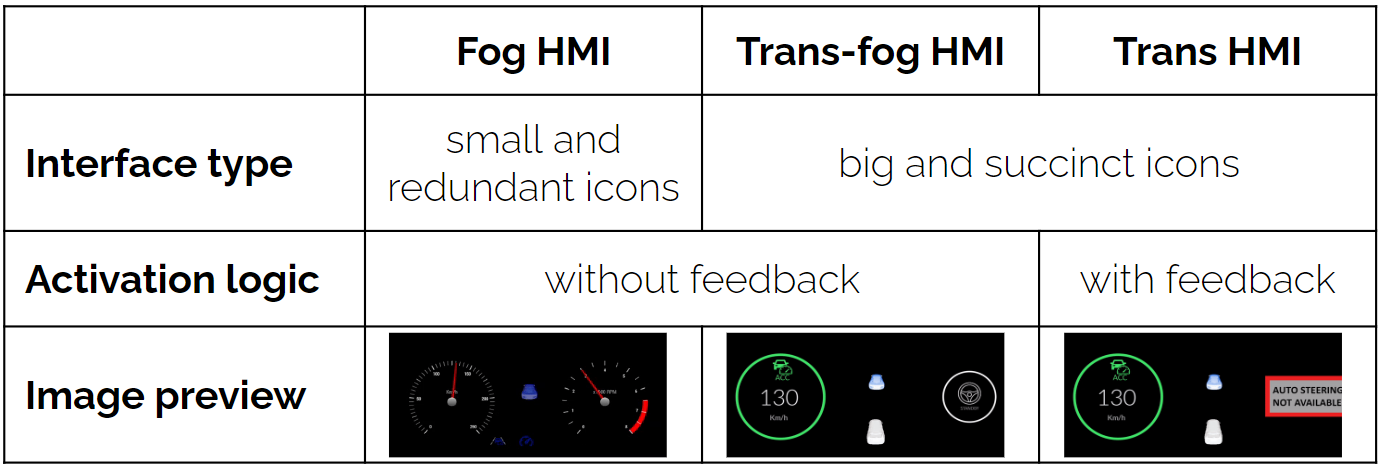}
     \caption{Illustration of HMI designs used in the driving simulator.}
     \label{fig:hmi}
\end{figure}

Three different SAE Level 2 HMI designs were developed and applied in the driving simulator, as shown in Fig.~\ref{fig:hmi}. Following the design principles for human-computer interface and results from the previous study \cite{liu2022transparency}, we developed these three in-vehicle HMIs that are distinct in transparency and workload required to understand. The \textit{Fog HMI design} should require the highest amount of mental workload during the interaction, owing to its small icons, low contrast color, and the fact that there would be no feedback when the system fails to activate. In contrast, the \textit{Trans HMI} provide clear graphical designs, with big and high contrast icon and feedback information when necessary, hence should provide participants with the highest transparency and relatively low mental workload during the automated driving. The last HMI design is the \textit{Trans-fog HMI}, which shares the same visual clarity as the \textit{Trans HMI} but has no feedback information, similar to the \textit{Fog HMI}.

In the previous study, these three HMI designs show significant differences in subjective workload measured (NASA-TLX). The same approach would also be included in this study as a reference to the previous one and others in the literature.

\subsection{Psychophysiological Measures}

The ECG electrodes were placed in the common Lead II Conﬁguration, while the EDA electrodes were mounted on the left foot to minimize the noise from the hand and right foot movement during driving. Usually, the EDA electrodes are placed on palmar sites or fingers, but foot placement is recommended when palmar sites or fingers are not available or suitable for placing EDA electrodes \cite{hossain2022comparison}. All electrodes are connected to a wireless trigger for LiveAmp, and the continuous signals were recorded and saved on a remote device throughout the experiment. The data were recorded with a 500 Hz sampling frequency and preprocessed in the Matlab version R2022a.

\begin{figure}[htbp]
     \centering
     \includegraphics[width=0.45\textwidth]{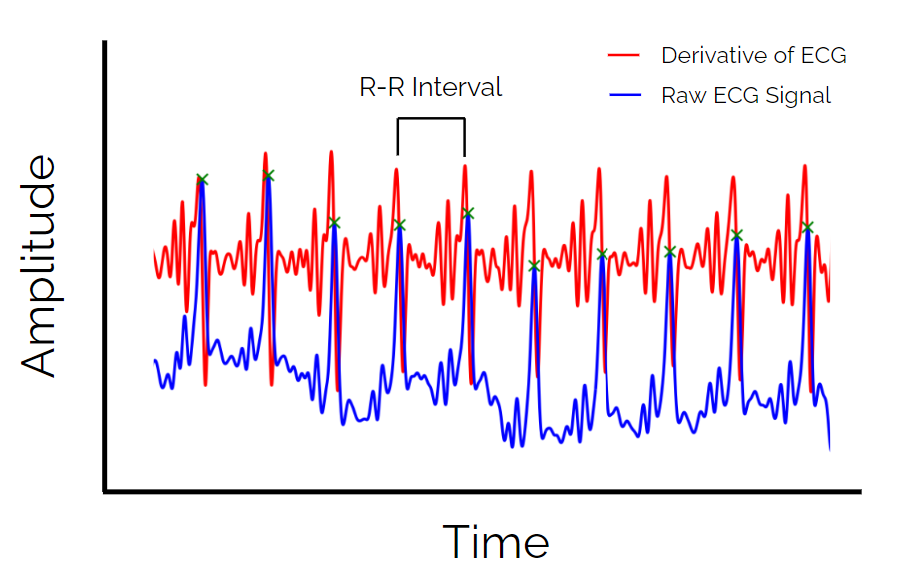}
     \caption{Illustration of the raw ECG signal and processed R-R interval.}
     \label{fig:ecg_signal}
\end{figure}

\subsubsection{Heart Rate Variability (HRV)}

In this study, we use the root mean square of successive differences between normal heartbeats (RMSSD) as the metric among the time domain heart rate variability (HRV) for its robustness and sensitivity in ultra-short-term workload measurement. To obtain the RMSSD, we first need to calculate the time differences between consecutive heart beats (or between R peaks). Then, average the squared values of those time differences. Finally, we take the square root of the average obtained and have the RMSSD over the designated duration. Before the RMSSD calculation, the raw ECG signal was filtered and differentiated for clearer R-R intervals, as shown in Fig.~\ref{fig:ecg_signal}. 
Ten seconds before and after the activation of the Level 2 ADS were used as an epoch representing the interaction period between the participant and the HMI design. During this time, participants were required to monitor the HMI closely to answer the following question when the driving simulator was paused or stopped.

\subsubsection{Skin Conductance Response (SCR)}

In this study, we intend to capture the rise of skin conductance With the EDA signal. As the cognitive load rises, so does the skin conductance value, and this change in EDA is the skin conductance response (SCR). After the initial stimulus, there is usually a 2-5 seconds delay before the skin conductance begins to rise. After the rising phase, the skin conductance value reaches the peak and begins to descend. This difference in skin conductance value between the initial stimulus and the peak after the delay is the SCR magnitude. We collected a 10 seconds epoch after the activation of the Level 2 ADS (stimulus) and calculated the corresponding SCR values.

\begin{figure}[htbp]
     \centering
     \includegraphics[width=0.45\textwidth]{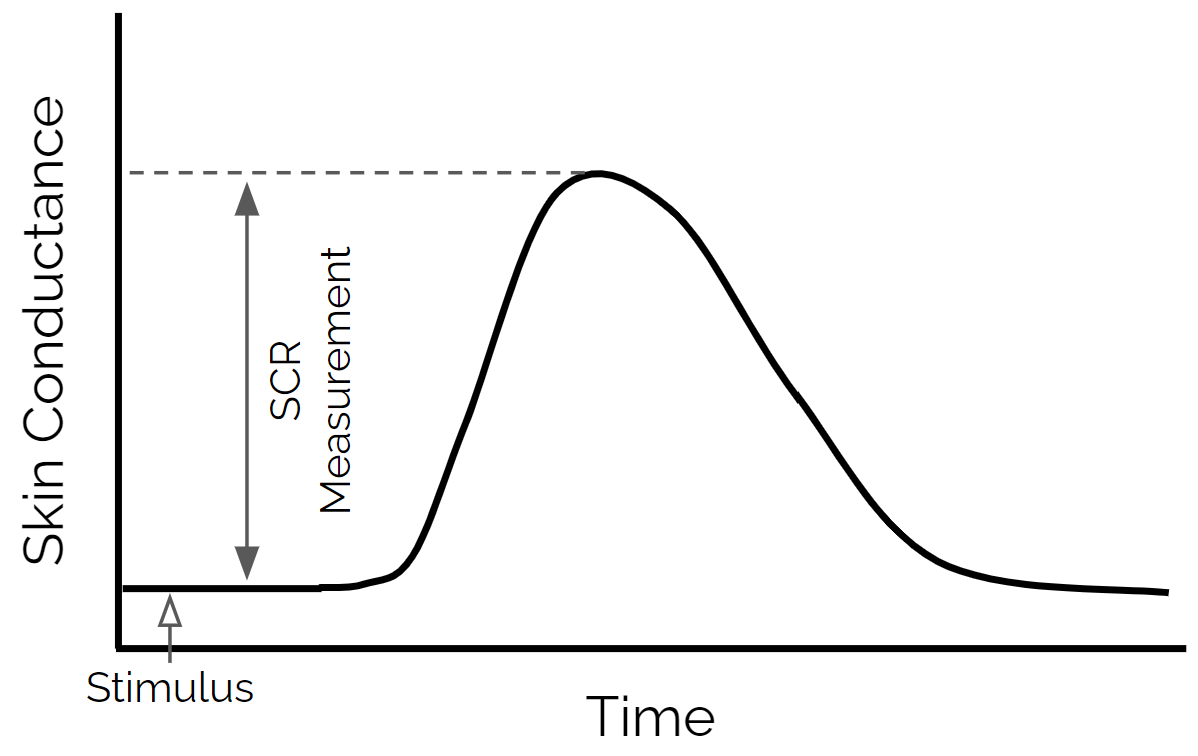}
     \caption{Illustration of the skin conductance response (SCR).}
     \label{fig:eda_signal}
\end{figure}

\subsection{Self-reported Workload Measures}

Two subjective workload measurements were included in this study. The first one is the National Aeronautics and Space Administration-Task Load Index (NASA-TLX) \cite{hart1988development}, which is a six-item questionnaire. The average across the six workload related factors was calculated without the weighted parameter. The second subjective workload measurement was derived from the previous study \cite{liu2022transparency}, where participants were asked to evaluate whether they agreed they could understand and obtain critical information from the HMI design. They were asked if they agreed that the HMI design was easy to understand. All three questions were scaled from 0 to 100.

\subsection{Procedure}

\begin{figure}[htbp]
     \centering
     \includegraphics[width=0.45\textwidth]{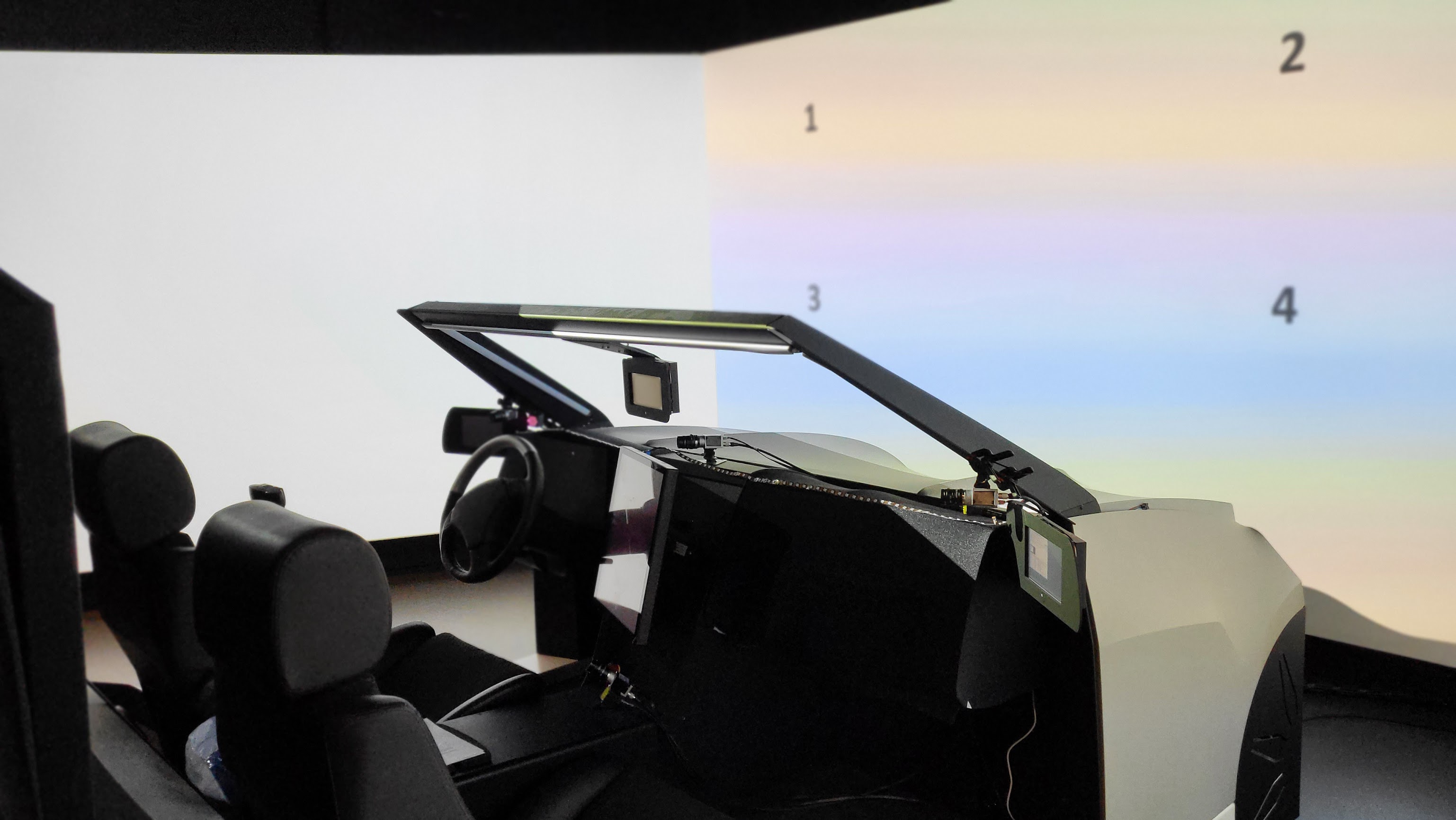}
     \caption{Illustration of the driving simulator setup.}
     \label{fig:simulator}
\end{figure}

The study was conducted in a static driving simulator with a field of view of 120$^{\circ}$, as illustrated in Fig.~\ref{fig:simulator}. The front panel consists of three screens with the scenes projected from three projectors respectively. It also has rear, left, and right mirrors, which are small LCD displays. The software SILAB was used to create the 4-lanes highway scenario. A touch screen was fixed on the right-hand side of the driver's seat, where automated cruise control (ACC) and L2 automation activation buttons were shown.

When participants arrived, we welcomed them and provided a brief study overview. A pre-recorded video containing detailed instructions was played to ensure all participants received the same information. Afterward, participants were asked to fill in a demographic questionnaire. In the meantime, experimenters set up the ECG and EDA electrodes and ensured signals from the amplifier were normal. Participants were then brought to the driving simulator and started a familiarization test drive. During the test drive, the baseline ECG and EDA data were also collected. After the familiarization, the formal test would begin if no symptoms of simulator sickness were shown.

\begin{figure}[htbp]
     \centering
     \includegraphics[width=0.45\textwidth]{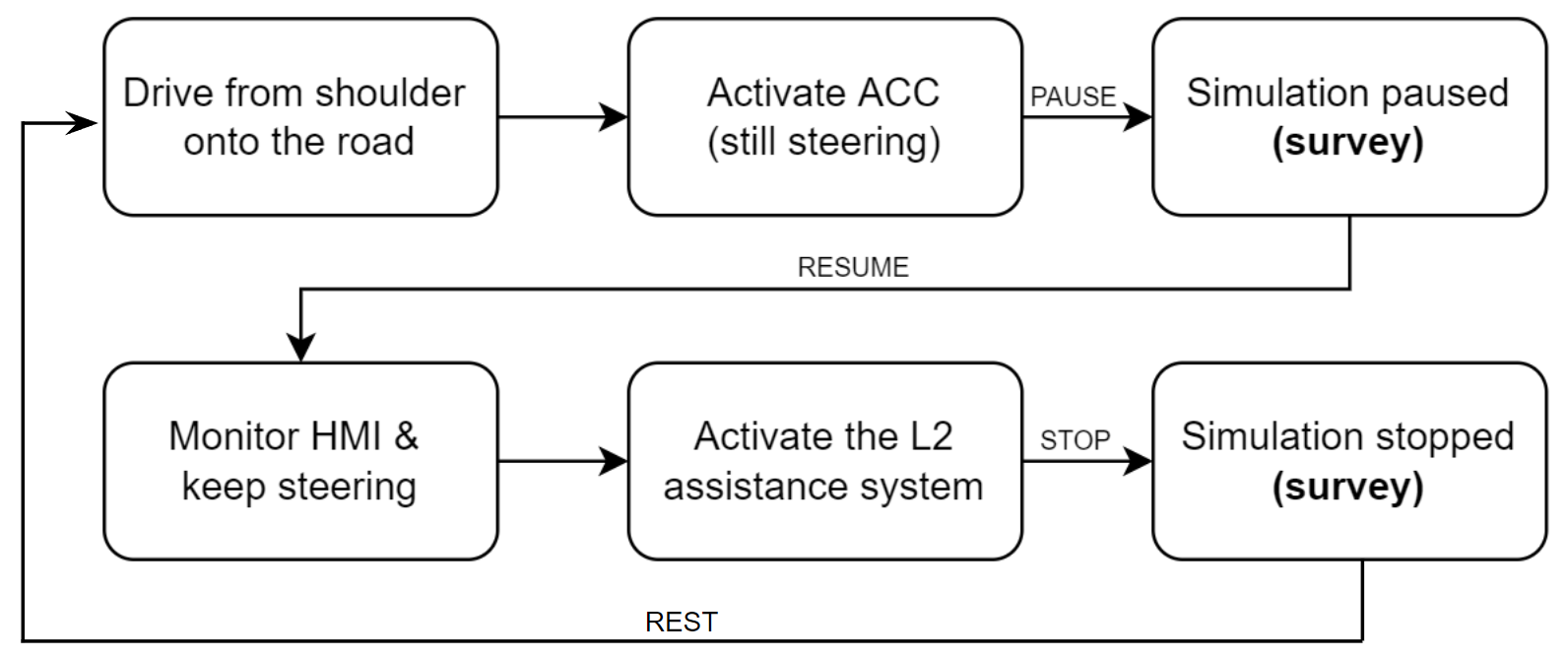}
     \caption{Experimental procedure.}
     \label{fig:procedure}
\end{figure}

The experimental procedure of this study is shown in Fig.~\ref{fig:procedure}. In each trial, one of the three HMI designs was applied randomly, and two surveys were presented to the participants at different stages. Both surveys consisted of a NASA-TLX questionnaire and a subjective transparency test and would later be used to compare to the psychophysiological measures. The first and second surveys were used to estimate the subjective workload the HMI design required when the ACC was activated, and when the Level 2 (L2) ADS was activated. respectively. In each trial, participants were asked to start the vehicle and drive from the shoulder to the right lane on the highway. Meanwhile, they could activate the ACC function on the panel right to the steering wheel whenever they felt comfortable. At the moment the ACC button was pressed, the ADS began the longitudinal control, followed by a five seconds countdown before the simulator went into a pause. When the simulator was paused, all the screens dimmed and the sound effects volume lowered in a fade-away pattern to avoid simulator sick. Then, participants were asked to complete the first survey and inform the experimenter when they finished. After the first survey, the simulation resumed, and participants were asked to activate the L2 ADS whenever they felt comfortable. Five deconds after the L2 button was pressed, the simulation stopped (also in a fade-away pattern), and the second survey was presented and asked to complete carefully. Participants had to go through all three HMI designs in a counterbalanced order, where for each HMI design there would be four different traffic layouts, making a total of 12 trials for each participant. The ECG and EDA signal was recorded throughout the study. Finally, participants were compensated at the end of the study after the feedback section about the procedure, questionnaires, and HMI designs.

\section{RESULTS}

In this study, we attempted to determine if different HMI designs would significantly affect the mental workload assessed with psychophysiological measures. However, self-reported workload measures were also used as a reference to the previous work and also to future studies. Repeated measure ANOVA was used to test the significance of the differences in ECG, EDA, NASA-TLX, and subjective transparency data. Multiple comparisons with Holm's correction were made if necessary.

\subsection{Psychophysiological Measures}
\begin{figure}
     \centering
     \begin{subfigure}[htbp]{0.239\textwidth}
         \centering
         \includegraphics[width=\textwidth]{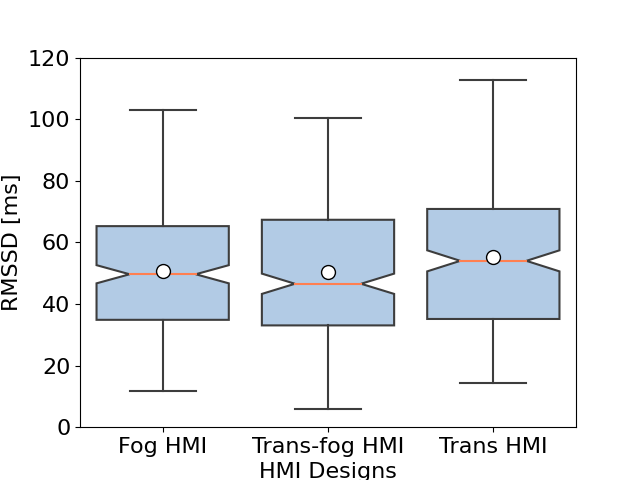}
         \caption{RMSSD.}
         \label{fig:RMSSD}
     \end{subfigure}
     \hfill
     \begin{subfigure}[htbp]{0.239\textwidth}
         \centering
         \includegraphics[width=\textwidth]{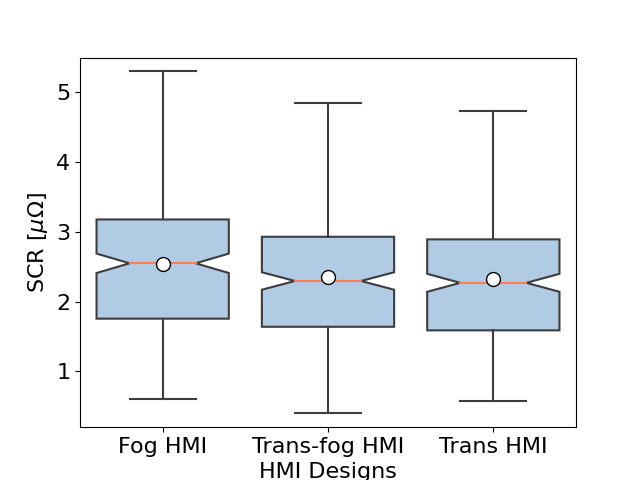}
         \caption{SCR.}
         \label{fig:SCR}
     \end{subfigure}
        \caption{Relationships between psychophysiological measures for workload and different HMI designs. (RMSSD: root mean square of successive differences between normal heartbeats. SCR: Skin conductance response)}
        \label{fig:psychophysio_hmi}
\end{figure}

We can see from Fig.~\ref{fig:psychophysio_hmi} that the same correlation in mental workload among HMI designs is shown, where the Trans HMI demanded the lowest workload during the interaction, followed by the Trans-fog HMI, and lastly, the Fog HMI. The time domain HRV, RMSSD, were found to be significantly different among those three HMI designs $F(2,522)=7.25, p<0.001, \eta_{p}^{2}=0.027$, where the Trans HMI had the highest RMSSD, representing that it demanded the lowest cognitive load during the interaction. A similar outcome was found on the SCR values, where the effect of HMI designs was also found to be significant $F(2,522)=3.372, p=0.035, \eta_{p}^{2}=0.013$, and that the Trans HMI had the lowest SCR, which again confirm the finding that the Trans HMI demand the lowest cognitive load when interacting with the ADS.

\begin{table}[htbp!]
\centering
\begin{threeparttable}
\caption{Post Hoc Comparisons on HMI Designs for Psychophysiological Measures.}
\label{table:psychophysio_posthoc}
\begin{tabular*}{0.45\textwidth}{@{\extracolsep{\fill}\quad}llcccc}
\toprule
\multicolumn{2}{c}{} & \multicolumn{2}{c} {\textbf{RMSSD}} & \multicolumn{2}{c}{\textbf{SCR}} \\
\cmidrule(rl){3-4} \cmidrule(rl){5-6}
  &   & $t$ & $p_{holm}$ & $t$ & $p_{holm}$ \\
\midrule
\makecell[c]{Fog} & Trans & -3.31 & \textbf{0.003} & 2.43 & \textbf{0.047}  \\
\makecell[c]  & Trans-fog & -0.024 & 0.98 & 2.09 & 0.075  \\
\makecell[c] {Trans} & Trans-fog & 3.29 & \textbf{0.003}  & -0.33 & 0.744 \\
\bottomrule
\end{tabular*}
\end{threeparttable}
\end{table}

As detailed in the Table.~\ref{table:psychophysio_posthoc}, post hoc comparisons were conducted for both psychophysiological measures. The RMSSD of the Trans HMI was found to be significantly higher than both the Fog and Trans-fog HMI designs, representing a lower workload demand than the rest two. No significant difference in RMSSD was found between the Fog and Trans-fog HMI designs. A similar result was found that the SCR of the Trans HMI design was significantly lower than that of the Fog HMI design. However, no significant difference was found between the Fog and the Trans-fog HMI designs and the Trans and the Trans-fog HMI designs.

\subsection{Self-reported Workload Measure}

\begin{figure}
     \centering
     \begin{subfigure}[htbp]{0.239\textwidth}
         \centering
         \includegraphics[width=\textwidth]{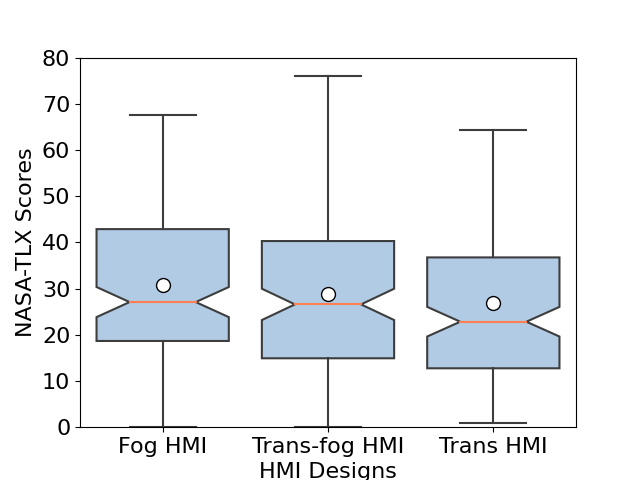}
         \caption{NASA-TLX.}
         \label{fig:nasatlx}
     \end{subfigure}
     \hfill
     \begin{subfigure}[htbp]{0.239\textwidth}
         \centering
         \includegraphics[width=\textwidth]{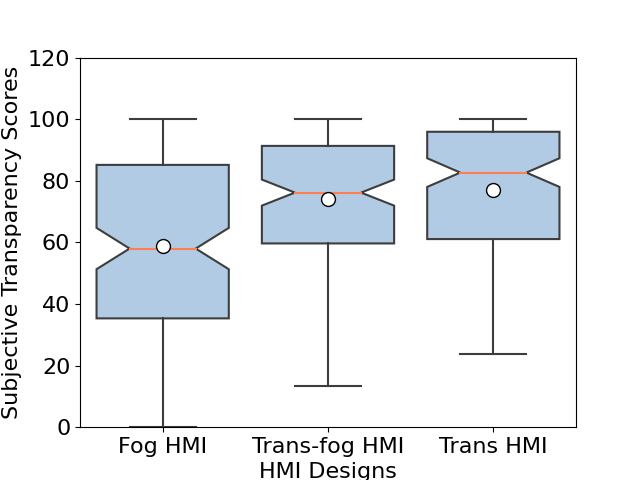}
         \caption{Subjective transparency test.}
         \label{fig:sub_trans_test}
     \end{subfigure}
        \caption{Relationships between self-reported measures for workload and different HMI designs.}
        \label{fig:self_report_hmi}
\end{figure}

We see the same pattern in self-reported mental workload among the HMI designs from Fig.~\ref{fig:self_report_hmi}, where the Trans HMI had the lowest subjective workload and highest self-reported transparency. The NASA-TLX scores were found to be significantly different among those three HMI designs $F(2,260)=3.80, p=0.028, \eta_{p}^{2}=0.027$, where the Trans HMI got the lowest averaged NASA-TLX score, representing that it required the lowest subjective mental workload. A similar outcome was found on the subjective transparency scores, where the effect of HMI designs was also found to be significant $F(2,260)=49.88, p<0.001, \eta_{p}^{2}=0.27$.

Post hoc comparisons were conducted for self-reported workload measures, as shown in Table.~\ref{table:self_report_posthoc}. The Trans HMI design had significantly lower NASA-TLX and higher subjective transparency scores than the Fog HMI design. But no significant difference was found between the Trans HMI and the Trans-fog HMI designs in NASA-TLX or subjective transparency scores. Between the Fog HMI and the Trans-fog HMI designs, the subjective transparency score of the Trans-fog HMI design was significantly higher than that of the Fog HMI design. But the effect was not found between these two in the NASA-TLX scores. 

\begin{table}[htbp!]
\centering
\begin{threeparttable}
\caption{Post Hoc Comparisons on HMI Designs for Self-reported Measures.}
\label{table:self_report_posthoc}
\begin{tabular*}{0.45\textwidth}{@{\extracolsep{\fill}\quad}llcccc}
\toprule
\multicolumn{2}{c}{} & \multicolumn{2}{c} {} & \multicolumn{2}{c}{\textbf{Subjective}} \\
\multicolumn{2}{c}{} & \multicolumn{2}{c} {\textbf{NASA-TLX}} & \multicolumn{2}{c}{\textbf{transparency}} \\
\cmidrule(rl){3-4} \cmidrule(rl){5-6}
  &   & $t$ & $p_{holm}$ & $t$ & $p_{holm}$ \\
\midrule
\makecell[c]{Fog} & Trans & 2.75 & \textbf{0.019} & -8.71 & \textbf{\textless{}0.001}  \\
\makecell[c]  & Trans-fog & 1.25 & 0.27 & -7.15 & \textbf{\textless{}0.001}  \\
\makecell[c] {Trans} & Trans-fog & -1,51 & 0.27  & 1.77 & 0.079 \\
\bottomrule
\end{tabular*}
\end{threeparttable}
\end{table}

\section{DISCUSSIONS}

We attempted to evaluate the sensitivity and effectiveness of psychophysiological measures in identifying different mental workloads required when interacting with different in-vehicle HMI designs in this study. We used the three HMI designs developed in the previous study, which were found to result in different self-reported workloads as the workload variable. Results from RMSSD suggested that the ECG measurement and the HRV analysis can be used to find differences in mental workload effectively. The Trans HMI was found to have the highest RMSSD during the interaction with participants, which suggests that it required the lowest workload among the three HMI designs. This was in accordance with the HMI designs and the other psychophysiological measures. The SCR was found to be the lowest on the Trans HMI design, representing that the Trans HMI design demanded the lowest workload since the rise of the SCR corresponded to the increase of the workload. Hence, the EDA, together with the SCR, could also be applied to investigate the differences in workload among HMI designs.

The self-reported workload measures were also confirmed to have the same correlation in workload with the previous study and psychophysiological measures, where the Trans HMI design had higher NASA-TLX scores than the Trans-fog HMI design, and the Trans-fog HMI design than the Fog HMI design. It was also not surprising that the Trans HMI design had the highest score in the subjective transparency test since the lower the workload needed to understand the HMI design, the higher the transparency should be. 

Traditionally, the HMI evaluation processes are usually subjective and heuristic, making it difficult to be efficient and standardized. Hence, to develop a standardized HMI evaluation method, the inclusion of objective measures is critical and necessary. In our previous work, we developed a transparency assessment method, which evaluates HMI transparency by combining the true understanding of the HMI and the workload required during the interaction. Now with the results from this study, the mental workload during the interaction between the users and the HMI designs could be estimated continuously and efficiently. Combining that with the proposed transparency assessment method, the assessment of the HMI designs could be adapted to environments with more dynamic interactions and be more efficient and reliable. 

\section{CONCLUSIONS AND FUTURE WORKS}
In this study, we confirmed the effectiveness of two psychophysiological measures in evaluating the mental workload when interacting with in-vehicle HMI design. This finding could be a strong basis for the HMI evaluation process. The next step is to include the psychophysiological measure in the proposed transparency assessment method, to develop an objective and standardized HMI assessment method, and use it to increase the efficiency of the HMI design process.

\addtolength{\textheight}{0cm}   




\section*{ACKNOWLEDGMENT}

This project has received funding from the European Union’s Horizon 2020 research and innovation program under the Marie Skłodowska-Curie grant agreement No. 860410.


\bibliographystyle{IEEEtran}
\bibliography{IEEEexample}

\end{document}